\documentstyle[aps,preprint]{revtex} 
\begin{document} 
\draft 
\tightenlines 
  
\title{Path-integral representation for a stochastic sandpile} 
\author{Ronald Dickman$^\dagger$ and Ronaldo Vidigal} 
\address{ 
Departamento de F\'{\i}sica, ICEx, 
Universidade Federal de Minas Gerais,\\ 
30123-970 
Belo Horizonte - MG, Brasil\\ 
} 
 
\date{\today} 
 
\maketitle 
\begin{abstract} 
We introduce an operator description for 
a stochastic
sandpile model with a conserved particle density,
and develop
a path-integral representation for its evolution.
The resulting (exact) expression for the effective action
highlights certain interesting features of the model,
for example, that it is nominally massless, and that
the dynamics is via cooperative diffusion.  Using the
path-integral formalism, we construct a diagrammatic 
perturbation theory,
yielding a series expansion for the activity density 
in powers of the time.
\end{abstract} 
\vspace{1cm} 


Short title: Path-integral for sandpile

\noindent {\small $^\dagger$electronic address: dickman@fisica.ufmg.br}

\newpage
\section{Introduction}

Sandpile models were introduced some fifteen years ago as examples
of self-organized criticality (SOC), or scale-invariance in the
apparent absence of control parameters 
\cite{bak,dhar,ggrin,sornet,vzprl}.  Subsequently the appearance
of such ``spontaneous" criticality was shown to result from a control
mechanism that forces the system to a critical point marking a
phase transition to an absorbing state \cite{socbjp,cancun}.   
In fact, sandpiles with the same local dynamics as the original
``self-organized" versions, but with strictly conserved particle
density, $p$, exhibit an absorbing-state phase transition as $p$
is varied.  Thus the particle density is the temperature-like
control parameter for these models.

Sandpiles with strictly conserved particle density (so-called
{\it fixed-energy sandpiles} or FES \cite{dvz}), exhibit
absorbing-state phase transitions \cite{marro,hinrichsen,bjp00}, 
which have attracted much interest of late, in connection with  
epidemics \cite{harris}, catalysis \cite{zgb,marro5}, 
and the transition to turbulence \cite{pomeau,chate,bohr}.
While continuous phase transitions to an absorbing state fall
generically in the universality class of directed percolation (DP) 
\cite{janssen,grassberger}, numerical evidence indicates that
this is not so for sandpiles \cite{fes2d,manna1d,mnrst,mancam}.
The non-DP nature of the transition has been attributed to the
coupling of the order parameter (the density of active sites), to
a second, conserved field (the local particle density), which
relaxes diffusively in the presence of activity, and remains frozen in its
absence \cite{rossi}.
Independent of the connection with SOC, fixed-energy sandpiles
furnish intriguing examples of phase transitions far from equilibrium, 
incorporating cooperative relaxation in the form of activated
diffusion.
Until now, all quantitative results for FES have been numerical,
based either on simulations \cite{fes2d,rossi,manna1d,mnrst}, 
or on coherent-anomaly analysis of a
series of $n$-site cluster approximations \cite{mancam}.  Phenomenological
field theories for sandpiles have been proposed 
\cite{paczuski,socprl,granada},
but their analysis is far from straightforward.
It is therefore of great interest to develop theoretical approaches
for FES.

This paper is the first of a series analyzing
a stochastic sandpile using operator and path-integral methods.  In this
work we establish the path-integral representation, and use it to
develop a time-dependent perturbation
theory that yields an expansion for the activity density in powers of
time.   In subsequent work the series will be extended and analyzed.

Our analysis uses two principal tools. The first is an operator formalism
for Markov processes, of the kind developed by Doi \cite{doi},
and applied to various models exhibiting nonequilibrium phase
transitions \cite{jsp1,tdpprl,jsp2,rsa,redner}. 
The second is an exact mapping, due to Peliti, from the Markov process
to a path-integral representation \cite{peliti85,pert}.
This approach is often used to generate an effective action, 
which may then be analyzed using renormalization
group (RG) techniques \cite{peliti86,lee,wijlanddp}.  
In the present instance,
however, we use the formalism not as the basis for an RG analysis,
but to generate a series expansion for the order parameter.
The expansion variable is the time; the coefficients are
polynomials in the particle density, $p$.

The balance of the paper is organized as follows.
In Sec. II we define the model and the master equation  
governing its dynamics using an operator formalism, which
is then mapped to the path-integral representation.  Sec. III
develops the perturbative expansion of the activity density
in powers of time.  In Sec. IV, we study some general properties of
the diagrammatic expansion, allowing us to
simplify and extend the analysis.  A brief comparison with
simulation results is presented in Sec. V.
A summary and discussion of our
results is given in Sec. VI.
 
\section{Evolution Operator} 
 
We consider Manna's stochastic sandpile in its fixed-energy 
(particle-conserving) version \cite{manna1d,manna,manna2}.   
The configuration is specified by the occupation number
$n_i$ at each site; sites with $n \geq 2$ are said to be 
{\it active}, and have a positive rate of
{\it toppling}.  When a site topples, it loses exactly two
particles (``grains of sand"), which move randomly and 
independently to nearest-neighbor (NN) sites.
In this work, we adopt a toppling rate of
$n(n\!-\!1)$ for a site having $n$ particles. 
This choice of rate represents a slight departure from the
examples studied previously, in which all active sites
have the same toppling rate.  The present rate leads to a much simpler
evolution operator, and, on the other hand, should yield the same
scaling properties, since sandpiles, like critical phenomena
in general, exhibit a high degree of universality. 
(Close to the critical point, the density of sites with $n \geq 3$
particles is quite low, so that in practical terms our choice of rate
should not alter quantitative properties greatly.
Since studies of {\it restricted-height} sandpiles \cite{mnrst,rossi}
reveal that they belong to the same universality class as their
unrestricted counterparts, there is good reason to expect that
a small change in transition rates, that does not modify the
symmetry or conservation laws of the model, will have no effect
on critical exponents.)   Preliminary simulation results \cite{rdunp}
indicate that the model studied here exhibits a continuous phase transition
at $p_c \!=\! 0.9493$, in one dimension; the corresponding value 
for the model with the same toppling rate for all active sites is
$p_c \!=\! 0.94885$ \cite{manna1d}.

For simplicity we begin the analysis in one dimension; the 
generalization to $d$ dimensions is straightforward. 
We represent the dynamics of this continuous-time Markov process
via the master equation, written in the form \cite{peliti85,pert}:
\[
\frac{d |\Psi \rangle}{dt} = L |\Psi \rangle ,
\]
where 
\[
|\Psi \rangle = \sum_{\{n_i\}} p(\{n_i\},t) |\{n_i\} \rangle
\]
is the probability distribution.
Here $p(\{n_i\},t)$ is the probability of the configuration
with occupation numbers $\{n_i\}$, and the state $|\{n_i\} \rangle$
is a direct product of states $|n_j \rangle$,
representing exactly $n_j$ particles at site $j$.

The evolution operator $L$ is written in terms of
creation and annihilation operators, defined via the relations:
\[
a_i |n_i\rangle = n_i |n_i\!-\!1\rangle
\]
and
\[
\pi_i |n_i\rangle = |n_i\!+\!1\rangle .
\]
The evolution operator for the one-dimensional stochastic sandpile
takes the form, 
 \begin{equation} 
L = \sum_i \left[ \frac{1}{4} (\pi_{i-1} + \pi_{i+1})^2 
 - \pi_i^2 \right] a_i^2 .
\label{evop} 
\end{equation} 
It is readily seen that $L$ conserves the number of particles.
 
It is convenient to work in the Fourier representation; 
on a lattice of $N$ sites with periodic boundaries, we introduce the 
discrete Fourier transform: 
 
\begin{equation} 
a_k = \sum_j e^{-ijk} a_j \;, 
\end{equation} 
with inverse 
\begin{equation} 
a_j = \frac{1}{N} \sum_k e^{ijk} a_k \;, 
\end{equation} 
(and similarly for $\pi_k$, etc.), 
where the allowed vales for the wave vector are: 
 
\begin{equation} 
k = -\pi, \; -\pi \!+\! \frac{2\pi}{N}, ...  
-\frac{2\pi}{N}, \; 0, \; 
\frac{2\pi}{N}, ..., \pi \!-\! \frac{2\pi}{N} \;. 
\label{brill} 
\end{equation} 
(To avoid heavy notation, we 
indicate the Fourier transform by the subscript $k$; 
the subscript $j$ denotes the corresponding variable on the lattice.) 
In Fourier representation, 
 
\begin{equation} 
L = -N^{-3} \sum_{k_1, k_2, k_3} \omega_{k_1, k_2} \pi_{k_1} 
     \pi_{k_2} a_{k_3} a_{-k_1-k_2-k_3} \;, 
\label{Lk} 
\end{equation} 
where $\omega_{k_1, k_2} = 1-\cos k_1 \cos k_2$. 
 
Since $L$ is in normal form (all operators $\pi$ to the left of all
operators $a$),
we can immediately write the 
evolution kernel, following Peliti's prescription \cite{peliti85,pert}: 
 
\begin{eqnarray} 
U_t(\{z_k\},\{\zeta_k\}) &=&  \int  \!{\cal D}  
\hat{\psi}  {\cal D} \psi 
\exp \!\left[\! -\!\int_0^t\! \!dt' \! \left\{ N^{-1}\sum_k 
\hat{\psi}_k \dot{\psi}_{-k}  \right. \right. 
 \nonumber \\ 
&+& N^{-3} \left. \left. \! \sum_{k_1, k_2, k_3} \omega_{k_1, k_2}  
\hat{\psi}_{k_1} \hat{\psi}_{k_2} \psi_{k_3} \psi_{-k_1-k_2-k_3} \right\}   
+ N^{-1} \sum_k z_k \psi_{-k}(t) \right], 
\label{ut} 
\end{eqnarray} 
with the boundary conditions $\psi_k(0) = \zeta_k$ and  
$\hat{\psi}_k(t) = z_k$.  (The dot denotes a time derivative; 
$\{z\}$ stands for the set of parameters $z_k$ 
associated with each wave vector, and similarly for $\{\zeta\}$.) 
The functional integrals are over the variables 
$\psi_k$ and $\hat{\psi}_k$: 
 
\begin{equation} 
\int  \!{\cal D} \hat{\psi} {\cal D} \psi  
\equiv \prod_k \int  \!{\cal D} \! 
\hat{\psi}_k \! \int \! {\cal D} \psi_k 
\end{equation} 
[The product is over the first Brillouin zone, Eq.(\ref{brill}).]
In the $d$-dimensional case, Eq. (\ref{ut}) remains valid if 
we let
\begin{equation}
\omega_{\bf k_1, k_2} = 1 - \lambda_d ({\bf k_1}) \lambda_d ({\bf k_1}) ,
\label{omegadd}
\end{equation}
with
\begin{equation}
\lambda_d ({\bf k}) \equiv \frac{1}{d} \sum_{\alpha = 1}^d \cos k_\alpha .
\label{deflamb}
\end{equation}
The wave vectors now range over the first Brillouin zone in $d$ dimensions.

The kernel is used to evolve the probability generating 
function,
$\Phi_t(\{z_j\}) \equiv \sum_{\{n_j\}} p(\{n_j\},t) \prod_j z_j^{n_j}$,
or its Fourier equivalent,
$\Phi_t(\{z_k\}) \equiv \sum_{\{n_k\}} p(\{n_k\},t) \prod_k z_k^{n_k}$,
via \cite{peliti85},
\begin{equation} 
\Phi_t(\{z_k\}) = \prod_k \int \frac{d\zeta_k d\zeta_k'}{2\pi}  
e^{-i \sum_k \zeta_k \zeta_{-k}'} 
U_t(\{z_k\},\{\zeta_k\}) \Phi_0(\{i\zeta_k'\})  \;\;\;(t\geq 0) . 
\label{map} 
\end{equation} 
We consider an initial product-Poisson distribution, 
with each site having the same mean number of particles, $p$, 
so that,  
 
\begin{equation} 
\Phi_0(\{z_j\}) = \exp[p\sum_j (z_j-1)] \;, 
\label{genfct0} 
\end{equation} 
corresponding to 
\begin{equation} 
\Phi_0(\{z_k\}) = \exp[Np (z_{k=0}-1)] \;, 
\label{genfct0k} 
\end{equation} 
in $k$-space. 
Noting that 
 
\begin{equation} 
\int \frac{d\zeta d\zeta'}{2\pi} f(\zeta) e^{i\zeta'(p- \zeta)} 
=  f(p) \;, 
\end{equation} 
we see that for a Poisson initial distribution,
 
\begin{equation} 
\Phi_t(\{z\}) = e^{-Np} U_t(\{z\};\zeta_k\!=\!Np\delta_{k,0}) \;. 
\label{genfctt} 
\end{equation} 
Thus the generating function at time $t$ is $e^{-Np}$ times the 
r.h.s. of Eq. (\ref{ut}), evaluated with  
$\psi_k(0) = Np \delta_{k,0}$. 
 
The simplest observables of interest are the mean particle number 
(at site $j$), 
 
\begin{equation} 
\langle n_j \rangle = \left. \frac{\partial \Phi_t(\{z_n\})} 
{\partial z_j} \right|_{z_n = 1}    \;, 
\label{den} 
\end{equation} 
and the mean activity, 
 
\begin{equation} 
\langle n_j (n_j\!-\!1) \rangle = \left. \frac{\partial^2 \Phi_t(\{z_n\})} 
{\partial z_j^2} \right|_{z_n = 1}    \;. 
\label{act} 
\end{equation} 
In $k$-space the condition $z_j\!=\!1$ for all sites $j$ becomes 
$z_k \!=\! N\delta_{k,0}$. 
Since the system is translation-invariant it is more convenient to 
study $\phi \equiv N^{-1} \sum_j \langle n_j \rangle$ 
and $\rho \equiv N^{-1} \sum_j \langle n_j (n_j-1) \rangle$. 
(Note that $\phi \!=\! p$ is a constant of the motion, so this 
quantity serves principally as a check on our analysis.) 
 
Let 
 
\begin{equation} 
U_t(\{z\},\{\zeta\}) =  \int  \!{\cal D}  
\hat{\psi}  {\cal D} \psi \; {\cal G}' [\psi,\hat{\psi}]  \;, 
\end{equation} 
where ${\cal G}'$ represents the exponential in Eq.(\ref{ut}), and let 
 
\begin{equation} 
\langle {\cal A} \rangle \equiv e^{-Np}   
\int  \!{\cal D} \hat{\psi}  {\cal D} \psi   
\left. {\cal A} \; {\cal G}' [\psi,\hat{\psi}]  
\right|_{z_k \!=\! N\delta_{k,0}} \;, 
\label{avg} 
\end{equation} 
where ${\cal A}$ is a function of the fields $\psi$ and $\hat{\psi}$. 
Evaluating the derivatives in Eqs. (\ref{den}) and (\ref{act}), we have 
 
\begin{equation} 
\phi = N^{-1} \langle \psi_{k=0} \rangle 
\end{equation} 
and 
 
\begin{equation} 
\rho = N^{-2} \sum_k \langle \psi_k \psi_{-k} \rangle  \;. 
\end{equation} 
 
It is convenient at this point to perform a change of 
variables, letting $\tilde{\psi}_k = \hat{\psi}_k - N\delta_{k,0}$. 
As a result, ${\cal G}'$ gains a factor of $e^{Np}$ [which cancels 
the prefactor in Eq. (\ref{avg})], and the boundary term in the 
argument of the exponential vanishes when we set  
$z_k \!=\! N\delta_{k,0}$.  Then we have 
 
\begin{equation} 
\langle {\cal A} \rangle =  \int  \!{\cal D}  
\tilde{\psi}  {\cal D} \psi   
\; {\cal A}\; {\cal G} [\psi,\tilde{\psi}] \;, 
\label{avga} 
\end{equation} 
where 
 
\begin{equation} 
{\cal G} [\psi,\tilde{\psi}] 
\equiv \exp \left[-N^{-1} \int_0^t dt' \sum_k \tilde{\psi}_k \dot{\psi}_{-k} 
+ \int_0^t dt' {\cal L}_I \right]  \;, 
\label{G} 
\end{equation} 
and where the ``interaction" is now: 
\begin{eqnarray} 
\nonumber 
{\cal L}_I &=&  
- N^{-3} \sum_{k_1, k_2, k_3} \omega_{k_1, k_2}  
\tilde{\psi}_{k_1} \tilde{\psi}_{k_2} \psi_{k_3} \psi_{-k_1-k_2-k_3}    
\\  
&-& 2N^{-2} \sum_{k_1, k_2} \omega_{k_1, 0}  
\tilde{\psi}_{k_1}  \psi_{k_2} \psi_{-k_1-k_2} \;. 
\label{LI} 
\end{eqnarray} 
Eq. (\ref{G}) with ${\cal L}_I \equiv 0$ defines ${\cal G}_0$; 
Eq. (\ref{avga}) with ${\cal G}_0$ in place of ${\cal G}$
defines $\langle {\cal A} \rangle_0$. 
 
The usual procedure at this point would be to take the 
continuum (small-$k$) limit,
generating a field theory for the process.  While this is not our purpose
in the present work, we note some interesting features of the model in this
context.  The first is that the interaction $L$ {\it contributes nothing}
to the quadratic part of the action.  (This can be seen immediately from
Eq.(\ref{evop}): $L$ is quartic in the fields.)  Thus the
resulting theory is nominally massless, and has {\it no evolution
at all} in the Gaussian approximation.  (Diffusion, in this model, is
cooperative, requiring the presence of at least two particles at the
same site.)
The action, moreover, contains no parameters whatsoever; the relevant
parameter $p$ is ``hidden" in the initial probability distribution.
A further difference from continuum descriptions of more familiar
processes such as DP \cite{janssen,grassberger} is that the order
parameter is given by $\langle \psi^2 \rangle$ not 
$\langle \psi \rangle$.

In simulations and cluster approximations 
\cite{fes2d,manna1d,mnrst,mancam}, FES clearly show a continuous 
phase transition between an active and an absorbing state as the 
parameter $p$ is varied, in close analogy to
more familiar examples, such as directed percolation.  
Thus at some more reduced level we might expect
to find an {\it effective} field theory of the usual kind, with a 
nonzero mass, bare diffusion coefficient, and one or more relevant 
parameters.  Such theories have indeed been proposed for
sandpile models \cite{socprl,granada}.  They include a second
field (the particle density) whose evolution is coupled to that
of the order parameter.  We leave the systematic
derivation of such a description, starting from the present exact
action, as a topic for future work.

\section{Perturbation Theory} 
 
\subsection{Free expectations}

To evaluate expectations of the form of Eq. (\ref{avga}) we write 
 
\begin{equation} 
\langle {\cal A} \rangle =  \langle {\cal A} e^{\int_0^t dt'{\cal L}_I}  
\rangle_0 \;, 
\label{avgb} 
\end{equation} 
with the intention of expanding the exponential.  Each term in the 
expansion (corresponding to a diagram, as specified below), can 
be evaluated once we have determined the free expectations of products of 
fields, $\langle \psi_{k_1}(\tau_1) \cdots \psi_{k_m} (\tau_m) 
\tilde{\psi}_{q_1} (\kappa_1) \cdots \tilde{\psi}_{q_n} (\kappa_n) 
\rangle_0$.  As usual, this can be reduced to expectations of 
a single field and of pairs of fields.  To begin, consider 
 
\begin{equation} 
\langle \psi_k(s) \rangle_0 =  \int  \!{\cal D} \tilde{\psi}  {\cal D} \psi 
\; \psi_k(s)  \exp \left[ -N^{-1} \int_0^t dt' \sum_k \tilde{\psi}_k 
\dot{\psi}_{-k} \right]  \;. 
\label{exppsi} 
\end{equation} 
The integrals over $\tilde{\psi}_k$ force the condition 
$\dot{\psi}_k = 0$, so that $\psi_k (s) \!=\! \psi_k(0) \!=\!  
Np \delta_{k,0}$.  Evidently, the same reasoning implies that  
free expectation of a product of $n$ fields $\psi$ is $(Np)^n$ 
if all the wave vectors are zero, and is zero otherwise. 
(The fact that a single field has a nonzero expectation may appear unusual, 
but in fact will not cause any inconvenience.  It is possible to 
make a further change of variables to $\psi_k' = \psi_k - Np \delta_{k,0}$, 
at the cost of introducing four new terms in ${\cal L}_i$.) 
 
To evaluate free expectations involving $\tilde{\psi}$, we 
define an operator ${\cal K}_q(\tau)$ via the property: 
 
\begin{equation} 
{\cal K}_q(\tau) {\cal G}_0 = \tilde{\psi}_k(\tau) {\cal G}_0  \;. 
\label{defK} 
\end{equation} 
Integrating by parts in the exponential, we have 

\begin{eqnarray}
\nonumber 
\frac{\delta}{\delta \psi_{-k} (\tau) } {\cal G}_0 
&=& \frac{\delta}{\delta \psi_{-k} (\tau) } 
\exp \left[ N^{-1} \int_0^t dt' \sum_{k'}\psi_{k'}  
\partial_{t'} \tilde{\psi}_{-k'}  
+  p\tilde{\psi}_{k=0}(t\!=\!0) \right] 
\\
&=& N^{-1} \partial_\tau \tilde{\psi}_k {\cal G}_0\;. 
\end{eqnarray} 
Recalling that $\tilde{\psi}_k(t) \!=\! 0$, we have 
 
\begin{equation} 
{\cal K}_q(s) = -N \int_s^t d\tau \frac{\delta}{\delta \psi_{-k} (\tau) } \;. 
\label{expK} 
\end{equation} 
Thus, 
 
\begin{equation} 
\langle \tilde{\psi}_k (s) \rangle_0 = 
-N \int  \!{\cal D} \tilde{\psi}  {\cal D} \psi \int_s^t d\tau  
\; \frac{\delta}{\delta \psi_{-k} (\tau) } \; {\cal G}_0 = 0 \;, 
\label{exppsit} 
\end{equation} 
where the final result is obtained via functional integration by parts. 
In the same manner we find the basic contraction (free propagator): 
 
\begin{equation} 
\langle \psi_{k'} (u) \tilde{\psi}_k (s) \rangle_0 = 
-N \int  \!{\cal D} \tilde{\psi}  {\cal D} \psi \int_s^t d\tau 
\; \psi_{k'} (u)  \;
\frac{\delta}{\delta \psi_{-k} (\tau) } \; {\cal G}_0 =  
N \delta_{k',-k} \Theta (u\!-\!s) \;, 
\label{contr} 
\end{equation} 
where $\Theta$ denotes the step function.  As is usual in this formalism, 
$\Theta(0) = 0$ \cite{pert}.  The free expectation of $n$ fields $\tilde{\psi}$ and 
$n$ fields $\psi$ is given by the sum of all products of $n$ pairwise 
contractions.  In case there are $m>n$ fields $\psi$ there are additional 
factors of $Np \delta_{k_i,0}$ associated with each uncontracted $\psi$. 
 
\subsection{Perturbative expansion}

We now have in hand all the ingredients needed to expand expectations 
of the form of Eq. (\ref{avgb}).  The first and second terms in  
${\cal L}_I$, Eq. (\ref{LI}), correspond to vertices with four and three 
lines (a ``4-vertex" and a ``3-vertex", respectively).  
Each $\tilde{\psi}_k (\tau)$ 
must be contracted with a $\psi_{-k} (\tau') $, where $\tau' > \tau$. 
(The required factors of $\psi$ may come from 
other vertices or from ${\cal A}$.) 
We adopt a graphical notation in which fields $\psi$ ($\tilde{\psi}$) 
are represented by lines entering (leaving) a vertex.  All lines are 
oriented toward the left, the direction of increasing time. 
Fig. 1 shows the vertices associated with ${\cal L}_I$, and the nodes 
that represent $\phi$ and $\rho$.  We refer to the latter 
as ``sinks" since they have no outgoing lines. 
Uncontracted fields $\psi$ are called ``external lines".
 
When we expand the exponential in Eq. (\ref{avgb}), the $n$-th 
order term carries a factor $1/n!$, and there are $n$ time integrations,  
$\int dt_1 \cdots \int dt_n$, all over the interval $[0,t]$. 
We absorb the factor $1/n!$ by fixing the time-ordering 
$t \geq t_1 \geq t_2 \geq \cdots \geq t_n \geq 0$.  This imposes certain 
restrictions on diagram topology, since a field $\tilde{\psi}$ must always be 
contracted with a $\psi$ at a later time.  Once this ordering is imposed, 
the integrand has no further time dependence, and the time integrations yield 
$t^n/n!$.      
 
Before formulating general rules, we study a few examples.  Consider 
 
\begin{equation} 
\rho = N^{-2} \sum_k \langle \psi_k \psi_{-k} e^{\int_0^t dt' {\cal L}_I} 
\rangle_0 \;. 
\label{exprho} 
\end{equation} 
The zeroth-order term is simply: 
 
\begin{equation} 
N^{-2} \sum_k \langle \psi_k \psi_{-k} \rangle_0 = p^2 \;, 
\end{equation} 
i.e., the initial activity for the product-Poisson distribution.
At first order we have the diagrams (a) and (b) shown in Fig. 2. 
For diagram (a) there is a combinatorial factor of 2, since there are 
two ways to contract the lines exiting the vertex with the fields in 
$\rho$; the contribution from this diagram is 
 
\begin{equation} 
-2 \int_0^t d\tau N^{-1}p^2 \sum_k \omega_{k,k} = - p^2 t 
\end{equation} 
In graph (b) we see that the condition that uncontracted $\psi$ fields 
have $k\!=\!0$ forces the line exiting the vertex to have $k\!=\!0$ 
as well.  But there is then a factor of $\omega_{0,0}\!=\!0$ associated 
with this vertex, so the contribution due to graph (b) vanishes. 
In general, we can exclude diagrams in which the two lines entering a 
3-vertex have momenta that sum to zero.   
 
At order $t^2$ 
we have diagrams (c) --- (f) shown in Fig. 2; (d) and (f) vanish for 
the same reason as (b).  (From here on we simply ignore such 
diagrams.) 
The contribution of graph (c) is readily 
obtained: there is a combinatorial factor of 4; the time integrations 
yield $t^2/2$; there is a remaining factor of $N^{-2}$ times 
the square of the sum encountered in graph (a).  The result is $p^2t^2/2$. 
In graph (e) there is a combinatorial factor of 8; its contribution is 
 
\begin{equation} 
16 (t^2/2) p^3 N^{-1} \sum_k [1\!-\! \cos^2 k][1\!-\! \cos k] 
= 4 p^3 t^2 \;. 
\end{equation} 
 
It is easy to see that our expansion conserves the particle density. 
With a one-line sink in place of the two-line sink corresponding to 
$\rho$, there will never be enough factors of $\psi$ to contract 
with all the $\tilde{\psi}$ factors, if use only four-line vertices. 
On the other hand, if we contract the sink with a three-line vertex, 
there will be a factor of $\omega_{0,0}\!=\!0$.  Thus 
$\phi (t) = \phi (0) = p$. 
 
\subsection{Diagram rules}

In the analysis that follows it will be convenient to employ the Laplace 
transform; the factor $t^n/n!$ then becomes $1/s^{n+1}$, where $s$ is the 
transform variable. 
Each diagram in the series for $\tilde{\rho}(s)$
carries a factor of $N^{L - 3n_4 - 2n_3 - 2}$, where 
$L$ is the number of lines (including external lines), and $n_3$ 
and $n_4$ are the numbers of 3-vertices and 4-vertices, respectively. 
There is exactly one factor 
of $N^{-1}$ for each free wave vector sum; from here on, we 
simply associate such a factor with each sum. 
We can formalize the foregoing discussion into a set of rules for 
finding the $n$-th order contribution to the order parameter  
$\tilde{\rho}(s)$: 

\noindent 1) Draw all connected diagrams with $n$ vertices, and a 
two-line sink to the left of all vertices.   
It is permissible for 
a line entering a node to be uncontracted (each external  
line carries a factor $p$ and must have momentum zero), but each line exiting a 
vertex ($j$) must be contracted with a vertex ($i<j$) to the left;  
there is a factor of $ \delta_{k',-k}$  
associated with each internal line.   (Here $k$ is the wave vector of 
the line exiting vertex $j$ and $k'$ the wave vector entering vertex $i$.) 
Given the restriction noted above, that the sum of the momenta entering a
3-vertex cannot be zero, the rightmost vertex of any diagram must be a
4-vertex.  We refer to this (and any other) 4-vertex with two external
lines as a {\it source point}.
 
\noindent 2) For each graph there is an overall factor of $1/s$, and 
a combinatorial factor counting the number of ways of realizing 
the contractions.  In the series for $\rho$, the combinatorial 
factor is the product of a factor $C_V$, associated with the choice 
of lines at each vertex (for a fixed set of connections between 
vertices), and a factor $C_L$ giving the number of choices of,
and connections 
between vertices, consistent with a given diagram. 
It is easy to show that 
$C_V = 2^Q$, where $Q = 1 + n_3 + 2n_4 - \ell - f$, with 
$\ell$ the number of simple loops, and $f$ the number of  
source points.  By a {\it simple loop} we mean a pair of vertices
directly connected by two lines, as in Fig. 2 (a).
The connection factor $C_L$ is unity for diagrams 
of three or fewer vertices, but can take nontrivial values for 
$n \geq 4$.  Examples are discussed below. 
 
\noindent 3) Each 3-vertex carries a factor of  
$-2 s^{-1} \omega_{k,0} = -2 s^{-1}  [1\!-\!\cos k]$, 
where $k$ is the momentum leaving the vertex.  Each 
4-vertex carries a factor of  
$ - s^{-1} \omega_{k_1,k_2}$, 
where $k_1$ and $k_2$ are 
the momenta of the lines exiting the vertex.  
 
\noindent 4) After 
taking into account all of the $\delta$-functions associated with 
propagators, the remaining free wave vectors are summed 
over the first Brillouin zone, Eq. (\ref{brill}). 

We close this subsection with a discussion of the
combinatorial factor $C_L$.    
As noted, $C_L$ represents the number of distinct choices of
vertices, and of connections between vertices, consistent with
a given diagram topology.  Evidently, diagrams with $n$ vertices
arise when we expand the product ${\cal L}_{I,1},...,{\cal L}_{I,n}$;
by ``choice of vertices" we mean whether the 3-vertex, or the 4-vertex,
associated with each ${\cal L}_{I,i}$ is selected.  
The $n$-th vertex is, as noted, always a 4-vertex.
Of course, if
the remaining $n\!-\!1$ vertices of a diagram are all of the same kind, 
then only one choice exists.  In most cases, exchanging the 
positions a 3-vertex 
and a 4-vertex yields a different diagram, but this is not always so.
Consider, for example, diagram (a) shown in Fig. 3.  
Let us refer to the $i$-th vertex in a given
diagram (in order of 
decreasing time) as $V_i$, with $V_0$ denoting the sink. 
In this diagram $V_1$ must be a 3-vertex, and $V_4$ a 4-vertex,
but we are free to choose between $V_2$ and $V_3$ as the other 3-vertex. 
Thus this diagram appears {\it twice} 
in the expansion of the activity, so that $C_L \!=\! 2$ in this case.
(That $C_L \!=\! 2$, and not more, rests on the fact that, given the
choice of vertices, there is only one way of connecting them to
yield the desired topology.)

Next we consider the possibility of different connections
among a fixed set of vertices.
We use $[i,j]$ to denote  
a link between $V_i$ and $V_j$. 
Consider diagrams (b) and (c) of Fig. 3.
For diagram (b), the set of connections between 
vertices is unique: $[0,1]$, $[0,4]$, 
$[1,2]$, $[2,3]$, and $[3,4]$, so 
$C_L \!=\! 1$ for this diagram. 
For diagram (c), by contrast, there are several different sets 
of connections that yield the same topology.  Evidently, the links 
between $V_1$ and $V_0$, 
and between $V_4$ and $V_3$, are obrigatory.  But this leaves open 
the question of which vertex the other outgoing line of $V_4$ links to, 
and of which vertex is linked to the other line entering the sink. 
One readily verifies that the following possibilities exist: 
 
\noindent i) $[0,1]$, $[0,2]$, $[1,4]$, $[2,3]$, $[3,4]$; 
 
\noindent ii) $[0,1]$, $[0,2]$, $[1,3]$, $[2,4]$, $[3,4]$; 
 
\noindent iii) $[0,1]$, $[0,3]$, $[1,2]$, $[2,4]$, $[3,4]$. 
 
\noindent Thus $C_L \!=\! 3$ for diagram (c).  Consider, finally, diagram (d)
of Fig. 3.  If $V_1$ and $V_2$ are both 3-vertices, 
there are two possible sets of connections:
 
\noindent i) $[0,1]$, $[0,2]$, $[1,3]^2$, $[2,4]$, $[3,4]$; 
 
\noindent ii) $[0,1]$, $[0,2]$, $[2,3]^2$, $[1,4]$, $[3,4]$. 

\noindent In addition, this diagram can be realized (with a unique set of
connections), with $V_1$ and $V_3$ as the 3-vertices, so that
$C_L \!=\! 3$ in this case.
 
\subsection{Examples and some general results}

There are seven diagrams at third order (Fig. 4).  Diagram (a), 
which continues the series whose first two terms are (a) and (c) 
of Fig. 2, is readily shown to yield $-p^2/s^4$. 
Applying the rules to diagram (b) in Fig. 4, 
we find: 
\begin{equation} 
16(-2) N^{-2} p^3 \frac{1}{s^4} \sum_k [1\!-\!\cos^2 k] 
\sum_{k'} [1\!-\!\cos^2 k'] [1\!-\! \cos k'] 
= -\frac{8}{s^4} p^3  \;. 
\end{equation} 
The contribution of diagram (c) is identical. 
For diagram (d) we have 
\begin{equation} 
16(-4) N^{-1} p^4 \frac{1}{s^4} \sum_k [1\!-\!\cos^2 k] [1\!-\! \cos k]^2 
= - \frac{40}{s^4} p^4  \;. 
\end{equation} 
Diagram (e) makes the same contribution. 
Similar analysis yields $-18 p^3/s^4$ for diagram (f) and  
$- 32 p^3 /s^4$ for (g). 
Collecting results, we have  
 
\begin{equation} 
\rho = p^2 - p^2 t + \frac{p^2t^2}{2} (1\!+\!8p)  
- \frac{p^2 t^3}{6}  
\left[1 + 66p + 80 p^2 \right] + {\cal O}(t^4) \;. 
\end{equation} 
 
In evaluating wave vector sums the following observations are helpful.  First, 
note that any wave vector sum $N^{-1} \sum_k$ can be written as 
$(2\pi)^{-1} \int_{-\pi}^\pi dk $.  
The analysis is facilitated by use of the identity: 
\begin{equation} 
I_n \equiv \frac{1}{N} \sum_k (1 \!-\! \cos^2 k) [1\!-\!\cos k]^n = 
4\cdot2^n \frac{(2n\!+\!1)!!}{(2n\!+\!4)!!}   \;. 
\label{id1} 
\end{equation} 
[This can be shown by writing $I_n$ as an integral, and letting 
$u \!=\! 1\!-\!\cos k$ so that 
\[ 
I_n = \frac{1}{\pi} \int_0^\pi du u^{n+1/2} \sqrt{2\!-\!u} \;. 
\] 
Integrating by parts one readily finds that 
\[ 
I_n = 2 \frac{2n\!+\!1}{2n\!+\!4} I_{n-1}   \;, 
\] 
leading to Eq (\ref{id1}), since $I_0 = 1/2$. 
For convenience we note:
$I_1 = 1/2$; 
$I_2 = 5/8$; 
$I_3 = 7/8$; and
$I_4 = 21/16$.]   Useful identitites in the $d$-dimensional case are:
\[
\sum_{\bf k} \lambda_d^2 ({\bf k}) = \frac{1}{2d},
\]
\[
\sum_{\bf k} \lambda_d^4 ({\bf k}) = \frac{3(2d\!-\!1)}{8d^3},
\]
and
\[
\sum_{\bf k} \lambda_d ({\bf k}) \lambda_d ({\bf q-k}) = 
\frac{1}{2d} \lambda_d ({\bf q}).
\]

Two general properties of the expansion are readily verified.  First,
the leading term at each order is $p^2 (-t)^n/n!$.  This set of contributions
sums to $p^2 e^{-t}$, describing the relaxation of the initial density of  
active sites.  In $d$ dimensions this becomes $p^2 \exp [-(2d\!-\!1)t/d]$.
The decay rate, $(2d\!-\!1)/d$, represents twice the probability that the
two particles released by a toppling site move to distinct neighbors, the factor
of two representing the toppling rate for a site with exactly two particles.

A second observation is that the highest power of $p$ 
appearing at order
$t^n$ is $p^{n+1}$, coming from diagrams with one source point and $n\!-\!1$
external lines terminating at 3-vertices.  The coefficient of this term
can be found as follows.  Note that the relevant diagrams consist of
($n\!-\!1$) 3-vertices attached to the lines of the simplest graph
(Fig. 2a).  The number of realizations of such topologies (that is,
the sum of the $C_L$ over all diagrams of this kind), is given
by one half the number of distinct sequences 
of $r$ symbols $A$ and $n\!-\!r\!-\!1$
symbols $B$ ($A$ and $B$ corresponding to 3-vertices
inserted in one or the other line), summed over $r$ from $r\!=\!0$ to $n\!-\!1$.
(The factor of one half is required because a pair of sequences that merely
exchange $A$'s and $B$'s in fact represent the same diagram.)
The number of such realizations is: 
\[
\frac{1}{2} \sum_{r=0}^{n-1}
\left( \begin{array} {c}
 n\!-\!1 \\
 r
	\end{array} \right) 
=2^{n-2}.  
\]
Each diagram of this kind
has a combinatorial factor $C_V = 2^{n+1}$, and an additional factor
of $(-1)^n 2^{n-1} I_{n-1}$ coming from the 3-vertices and the 
wave vector sum.   Using our result for $I_n$, we find that this 
contribution is:
\[
(-1)^n 2^{4n-1} \frac{(2n-1)!!}{(2n+2)!!} \frac{p^{n+1} t^n}{n!},
\]
for $n \geq 2$.  Call the sum of such contributions $\rho_{max}(t)$;
it can be evaluated in closed form by noting that
\begin{eqnarray}
\nonumber
\rho_{max}(t) &=& \frac{p}{4} \frac{1}{2\pi} \int_{-\pi}^{\pi} dk
(1 \!-\! \cos^2 k) \sum_{n=2}^\infty 
\frac{(-8pt)^n(1\!-\! \cos k)^{n-1}}{n!}
\\
&=& \frac{p}{4} \left\{ e^{-8pt} \left[ I_0 (8pt) + I_1 (8pt) \right] 
-1  \right\} + p^2 t ,
\end{eqnarray}
where $I_\nu $ denotes the modified Bessel function.  The factors
of $e^{-8pt} I_\nu (8pt)$
imply slow relaxation ($\sim 1/\sqrt{t}$)
typical of diffusive processes.

\section{Diagrammatic analysis} 
 
As is evident from the foregoing discussion,
many diagrams represent simple variations of those appearing 
at some lower order.  The calculation will be simplfied if we can 
identify classes of related diagrams. 
To begin we enumerate some ways in which diagrams ($g'$) with $v+1$ vertices 
can be formed from a $v$-vertex diagram $g$.  
 
\noindent a) Insert a 3-vertex into any internal line; $g'$ has one more 
external line than does $g$.  An example is the generation of diagram (e), 
starting from (a) in Fig. 2. 
 
\noindent b) Replace any 4-vertex by a ``4-loop", consisting of a pair of 
4-vertices joined by two lines.  Diagram (c) (Fig. 2) is generated from 
(a) by this process. 
 
\noindent c) Replace any 3-vertex V by a 4-vertex, contracting the new 
outgoing line with a {\it new} 3-vertex 
inserted in an existing internal line.  In case the contraction is 
to a 3-vertex V$'$ {\it on the same line} as V, and there are {\it 
no other vertices on this line}, between V and V$'$, we call the  
resulting structure a ``3-loop".  Starting with diagram (e) in Fig. 2, 
we may generate (f) and (g) (Fig. 4) by this procedure. 
Diagram (f) has a 3-loop but (g) does not, since in this 
case V$'$ does not lie on the same line as as V. 
 
\noindent d) Insert a 4-vertex into an internal line, contracting one of its 
outgoing lines to an existing external line.  (The vertex associated 
with the latter must lie to the left of the new 4-vertex.)  Graphs 
(f) and (g) (Fig. 4) are formed from (e) (Fig. 2) by this means. 
 
\noindent e) Add a new 4-vertex to $g$ by joining its outgoing lines to two 
external lines in $g$.  Consider again (e) in Fig. 2; if we join two 
external lines (one of them associated with the 3-vertex) to a new 
4-vertex, the result (after some redrawing) is (g) of Fig. 4. 
 
Now consider the inverse of operations a) - c).  Given a graph $g$ 
with two or more vertices: 
 
\noindent a') Remove all 3-vertices bearing an external line.  
 
\noindent b') ``Collapse" any 4-loop to a single 4-vertex. 
 
\noindent c') Collapse any 3-loop to a 3-vertex. 
 
The systematic application of these steps, until no further subtractions  
are possible, will be called {\it reduction} of a diagram.  
Let us define an {\it articulation point} as a 4-vertex (other than a 
source point), which, if removed, 
will leave the diagram disconnected.   
(Such a vertex has, by necessity, total momentum zero entering it.) 
In Fig. 2 only graph (c) has 
an articulation point, while graphs (a), (b) and (c) of Fig. 4 all possess 
one or more such point.   
 
We may now define an 
{\it irreducible} diagram (IRD) as one free of 3-vertices bearing 
external lines, free of collapsible loops, and having no articulation 
points.  The simplest irreducible diagram is (a) of Fig. 2; 
we shall refer to it as the one-loop IRD.   
At each order only a small minority of the diagrams are irreducible. 
At second order there are {\it no} IRDs; the only one 
at third order is (g) in Fig. 4.
The IRDs with four vertices are shown in Fig. 5; there are 23 
IRDs with five vertices.
 
In addition to the local modifications (a - e) described above, 
we can also form new diagrams via {\it composition}. 
Given two diagrams $g$ and $g'$ we form the {\it composite diagram} 
$gg'$ by: 
 
\noindent i) Deleting the two lines entering one of the source points $r$ of $g$; 
 
\noindent ii) Identifying the sink in $g'$ with $r$ in $g$. 
 
\noindent (If $g$ and $g'$ are distinct we can also form $g'g$.)   
Graph (b) of Fig. 4 
is composed, in this sense, from (a) and (e) of Fig. 2. 
Composite graphs are reducible by definition (the two components are 
joined at an articulation point.)  When we have removed all 3-vertices 
with external lines, and collapsed all collapsible loops in a diagram, 
the result is either an IRD or a composite of 
such diagrams.

The expansion up to order $t^v$ may be organized as follows.  First 
we identify all of the IRDs with $v$ or fewer vertices.  For each such
IRD, we evaluate its contribution and that of all diagrams with
$\leq v$ vertices that are directly reducible to it.  There remain
the composite diagrams.
Let $\tilde{C}_g (s)$ be the contribution to $\tilde{\rho}(s)$ 
due to diagram $g$.  From the definition 
of the composition process, we see that the contribution of a 
composite diagram $gg'$ to the activity is: 
 
\begin{equation} 
\tilde{C}_{gg'} = \frac{s}{p^2} \tilde{C}_g \tilde{C}_{g'}. 
\label{ccomp} 
\end{equation} 
We make use of this relation as follows. 
For each IRD $g^*$, define 
 
\begin{equation} 
\tilde{F}_{g^*}(s) = \frac{s}{p^2} \left[ \tilde{C}_{g^*}(s) +  
\sum_{g' \succ g^*} \tilde{C}_{g'}(s) \right] \;, 
\end{equation} 
as $s/p^2$ times the sum of contributions due to $g^*$ {\it and all graphs 
reducible to it} (we use $g' \succ g^*$ to mean 
``$g'$ is reducible to $g^*$").   
Now define 
 
\begin{equation} 
\tilde{F} (s) = \sum_{g^*} \tilde{F}_{g^*}(s) \;, 
\end{equation} 
and 
\begin{equation} 
\tilde{G} (s) = \sum_{g^*} n_{g^*} 
\tilde{F}_{g^*}(s) \;, 
\end{equation} 
where the sum is over all IRDs, and 
and in the second sum $n_{g^*}$ is the number of  
source points in $g^*$.  Each 
source point serves as a possible attachment site for the preceding 
element.   Then the sum of (i) all irreducible diagrams, (ii) all 
diagrams reducible to an IRD, and (iii) all linear composite diagrams is 
 
\begin{equation} 
\tilde{\rho}_L (s) = \frac{p^2}{s}  
\frac{\tilde{F} (s)}{1 - \tilde{G} (s)} \;. 
\end{equation} 
Missing from $\tilde{\rho}_L$ are  
{\it branching} composite diagrams. 
These are composite diagrams in which at least  
one component has two or more diagrams attached to its 
source points.  Since diagrams with more than one source
point have at least four vertices, branching diagrams
must have at least six.  Thus $\tilde{\rho} (s) = \tilde{\rho}_L (s)$ 
to ${\cal O} (s^{-6})$.
Evaluating $\tilde{F}$ to and $\tilde{G}$ to this order, 
we obtain 

\begin{eqnarray}
\nonumber
\rho &=& p^2 - p^2 t + \frac{p^2t^2}{2} (1\!+\!8p)  
- \frac{p^2 t^3}{6}  
\left[1 + 66p + 80 p^2 \right] 
\\ \nonumber
&+& \frac{p^2 t^4}{4!}  
\left[1 + 442p + 2076 p^2 + 896p^3\right] 
\\ 
&-&\frac{p^2 t^5}{5!}  
\left[1 + 2842p + 35396 p^2 + 52240p^3 + 10752 p^4 \right] 
+ {\cal O}(t^6) \;. 
\label{rhot}
\end{eqnarray}
In $d$ dimensions the result is: 

\begin{eqnarray}
\nonumber
\rho &=& p^2 - \gamma p^2 t + \frac{p^2t^2}{2} \gamma (\gamma \!+\!8p)  
\\
\nonumber
&-& \frac{p^2 t^3}{6}\left[\gamma^3 
+ \frac{2p}{d^4} \left(80d^4 \!-\! 56 d^3 \!+\! 12 d^2 
\!-\! 6d \!+\! 3 \right)
+ \frac{16 p^2}{d^3} \left(8d^3 \!-\! 6d \!+\! 3 \right) \right] 
\\
\nonumber
&+& \frac{p^2 t^4}{4!} \left[ \gamma^4
+ \frac{p}{d^5} \left(1152 d^5 \!-\!960d^4 \!+\! 336d^3 \!-\!128d^2 
\!+\! 36d \!+\!6 \right) 
\right.
\\
\nonumber
&\;\;+& \left. \frac{4p^2}{d^5} \left(960d^5 \!-\! 368d^4 \!-\! 112d^3 
\!-\! 48d^2 \!+\! 108 d \!-\! 21 \right)
+ \frac{128p^3}{d^3} \left(8d^2(d\!+\!1) -9(2d\!-\!1) \right) \right]
\\
&+& {\cal O}(t^5) \;, 
\end{eqnarray}
where $\gamma \equiv (2d\!-\!1)/d$.

\section{Comparison with simulation}

Although the series we have derived are too short to yield reliable
predictions for the activity $\rho(t)$ at long times, it is of interest 
to consider a preliminary comparison with simulation results, as a
check on our analysis.  We have simulated the one-dimensional model on
lattices of $L=$1000 and 2000 sites, with periodic boundaries; initially
$N\!=\!pL$ particles are placed randomly on the lattice, generating 
a product-Poisson distribution of occupation numbers $n_j$.  We average
over $5 \times 10^5$ independent realizations of the process.

A typical result is shown in Fig. 6, for $p\!=\!1/2$.  Since this is well
below the critical value, $\rho(t) \to 0$ as $t \to \infty$.  (Note that
the average is over all realizations, including those that have
fallen into the absorbing state.)  The simulation result (data points
merging to the bold line) decays quite rapidly at first, and then
approaches zero more slowly; a systematic study \cite{rdunp} reveals
that the approach to $\rho \!=\! 0$ is best characterized as a
stretched exponential, $\rho \sim \exp[-at^\beta]$ with $\beta \simeq 1/2$.

Two theoretical curves derived using the series, Eq. (\ref{rhot}),
are shown in Fig. 6.  The upper one is generated by
transforming the time series to the variable $y = [1 \!-\! e^{-bt}]/b$,
and then constructing the [3,2] Pad\'e approximant to the $y$-series.
For $b$ in the range 0.25 to 0.4 we find good agreement at 
short times; for $b\!=\! 0.35$ (shown here) there is excellent agreement
for $t \leq 10$, but the theoretical prediction attains a constant,
nonzero value thereafter.  

Since $\rho(t)$ must decay to zero in this case, it is
interesting to study the series for $d \ln(\rho/p^2)/dt$.  The latter
remains negative for large $t$, so that the
activity itself decays to zero.  The lower curve in Fig. 6 is obtained by
transforming the derivative-log series to the variable 
$z = 1 - (1 \!+\!bt)^{-1/3}$ (with $b\!=\!2.7$), 
and forming the [2,2] approximant to the
$z$-series, which is then used to evaluate $\ln (\rho/p^2)$ via
numerical integration.  The activity does indeed decay to zero, but
too rapidly.  We obtain very similar results using $b$ in the range
of 2 to 3, and using other transformations, for example $y$ given above,
or a modified $z$ with the exponent 1/2 in place of 1/3.
After examining various further transformations and analyses,
we conclude that the present series is simply too short to yield
useful predictions for long times.

It should be noted
that the apparent radius of convergence of the original series
(estimated from ratios of successive coefficients), 
is about $t < 1/4$ for $p \!=\! 1/2$; evidently, the transformation and
approximant analysis greatly extends the utility of the series.
(For $p\!=\! 1$, the radius of convergence is about $t < 1/6$; the
transformed series agrees well with simulations for $t \geq 2$.)
Thus, despite the divergence between simulation and theory for times greater than
about 10, we regard the present comparison, based as it is on a very short
series, as confirming the validity of our analysis.  Since the approach
to the stationary state is neither exponential nor power-law, we
anticipate that a fair number of terms will be required to represent it.
Once extended series are available, transformations that are consistent
with the asymptotic behavior should yield good numerical estimates.

\section{Discussion} 
 
We have derived an exact path-integral representation for the
dynamics of a stochastic sandpile model similar to Manna's
sandpile, in a version with strictly conserved particle number.  
(The key differences are: (1) We analyze a continuous-time 
process, whereas Manna's model is defined in discrete time; and,
(2) The toppling rate for a site with $n$ particles is $n(n\!-\!1)$
rather than uniform, for all sites with $n \geq 2$.)  We have argued that
such minor differences should not affect scaling properties. 
The path-integral mapping yields an effective action that is
nominally massless and parameter-free, the relevant parameter $p$
(the particle density), appearing in the initial distribution
not the evolution operator.  This conclusion applies generally
to non-dissipative sandpiles.  

In contrast to phenomenological field theories of sandpiles
\cite{socprl,granada}, in which the order parameter density
is coupled to a second slowly relaxing variable (the particle 
density), the exact action derived in the present work involves
but a single field, $\psi$, whose expectation is the particle density.
The order parameter is given by $\langle \psi^2 \rangle$.
The derivation of a phenomenological
description along the lines of Refs. \cite{socprl} and \cite{granada},
starting from the exact action, remains an open question.

We develop a diagrammatic perturbation theory, leading to 
an expansion for the activity density in powers of time,
where the coefficients are polynomials in $p$.  We point out certain
general properties of the expansion: its limiting forms 
for very small, and very large $p$, and the slow, diffusion-like
nature of the relaxation.  The series provides an important check
on other theoretical methods, and on simulation results.
While it appears to have a limited radius of convergence, experience
with similar series, for the contact process \cite{jsp2} and for
random sequential adsorption \cite{baram,rsa}, indicates that under
suitable transformation of variables and Pad\'e approximant
analysis, useful predictions can be obtained.
We present some preliminary numerical evidence
for this assertion.
Thus, if the series can be extended, it should
be possible to investigate scaling properties of the sandpile.
We intend to pursue such extensions and analyses in the near future.

\newpage
\noindent FIGURE CAPTIONS 
\vspace{1cm} 
 
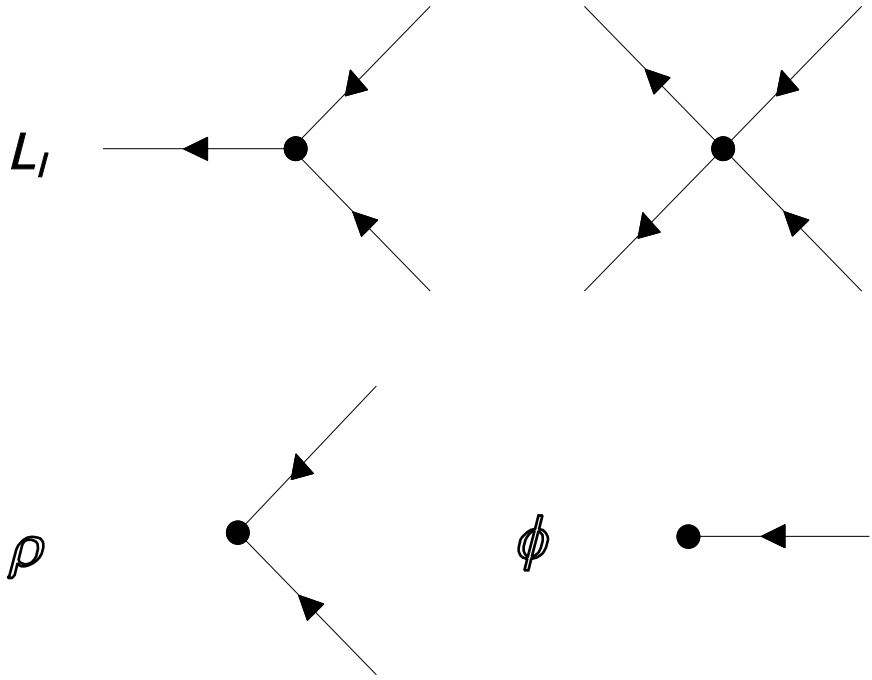
\noindent Fig. 1. Vertices and sinks in ${\cal L}_I$, $\rho$,  
and $\phi$. 
\vspace{1em} 
 
\noindent Fig. 2. One- and two-vertex diagrams in the expansion of the  
order parameter. 
\vspace{1em} 
 
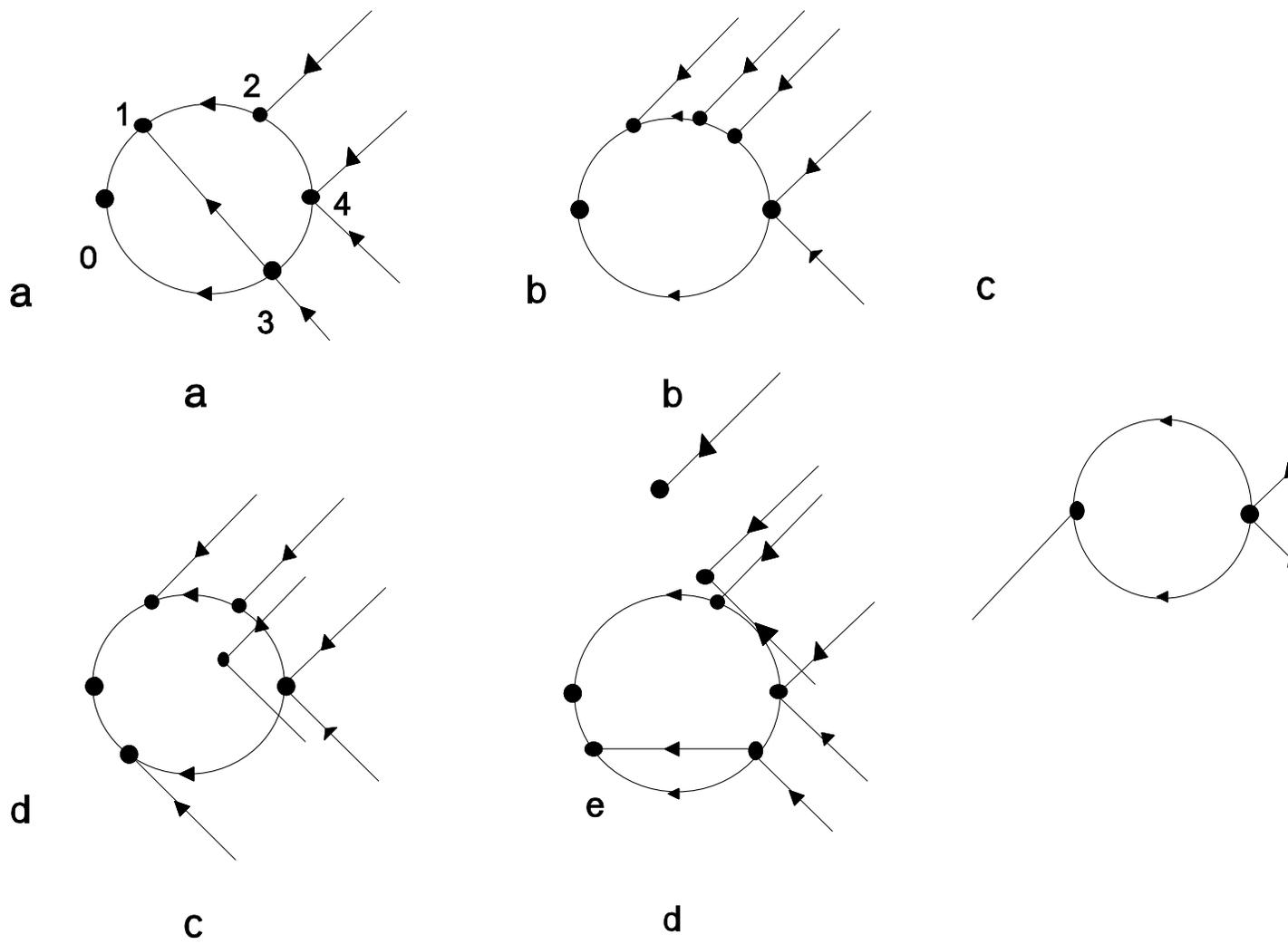
\noindent Fig. 3. Diagrams analyzed in the discussion in the combinatorial
factor $C_L$. 
\vspace{1em} 

\noindent Fig. 4. Diagrams with three vertices.  Only (g) is irreducible. 
\vspace{1em}

\noindent Fig. 5. Irreducible diagrams with four vertices. 
\vspace{1em} 
 
\noindent Fig. 6. Activity in the sandpile with $p \!=\! 1/2$. 
The points (merging to a bold line) are simulation results;
the curves above and below are series-based predictions discussed in the
text.

\begin{thebibliography}{99} 

\bibitem{bak}
	P. Bak, C. Tang and K. Wiesenfeld,
	Phys. Rev. Lett. {\bf 59}, 381 (1987);
	Phys. Rev. A {\bf 38}, 364 (1988).

\bibitem{dhar}
	D. Dhar, 
	Physica A {\bf 263} (1999) 4, and references therein.

\bibitem{ggrin}
	G. Grinstein, in 
	{\it Scale Invariance, Interfaces and Nonequilibrium Dynamics}, 
	{\it NATO Advanced Study Institute, Series B: Physics},
	vol. 344, A. McKane et al., Eds. 
	(Plenum, New York, 1995).

\bibitem{sornet}
	D. Sornette, A. Johansen, and I. Dornic,
	J. Phys. I (France) {\bf 5}, 325 (1995).        

\bibitem{vzprl}
	A. Vespignani and S. Zapperi,
	Phys. Rev. Lett. {\bf 78}, 4793 (1997);
	Phys. Rev. E {\bf 57}, 6345  (1998).

\bibitem{socbjp}
	R. Dickman, M. A. Mu\~noz, A. Vespignani, and S. Zapperi,
	Braz. J. Phys. {\bf 30}, 27 (2000).
	e-print: cond-mat/9910454.

\bibitem{cancun}
	    R. Dickman, 
	    Physica A{\bf 306}, 90 (2002).

\bibitem{dvz}
	R. Dickman, A. Vespignani and S. Zapperi,
	Phys. Rev. E {\bf 57}, 5095 (1998).

\bibitem{marro}
	  J. Marro and R. Dickman 
	  {\em Nonequilibrium Phase Transitions in Lattice Models} 
	  (Cambridge University Press, Cambridge, 1999).

\bibitem{hinrichsen}
	     H. Hinrichsen, Adv. Phys. {\bf 49}, 815 (2000).


\bibitem{bjp00} 
	 Various articles on absorbing-state phase transitions are collected in
	 Braz. J. Phys. {\bf 30}, (2000).

\bibitem{harris}
	 T. E. Harris, 
	 Ann. Prob. {\bf 2} (1974) 969.

\bibitem{zgb}
	 R. M. Ziff, E. Gulari, and Y. Barshad,
	 Phys. Rev. Lett. {\bf 56} (1986) 2553.

\bibitem{marro5}
	 See Ch. 5 of Ref. \cite{marro}.

\bibitem{pomeau}
	 Y. Pomeau,
	 Physica D{\bf 23} (1986) 3.

\bibitem{chate}
	 H. Chat\'e and P. Manville, 
	 Phys. Rev. Lett. {\bf 58}, 112 (1986).

\bibitem{bohr}
	 T. Bohr, M. van Hecke, R. Mikkelsen, and M. Ipsen,
	 Phys. Rev. Lett.  86 (2001) 5482.

\bibitem{janssen}
	  H. K. Janssen, Z. Phys. B 42 (1981) 151.

\bibitem{grassberger}
	    P. Grassberger, 
	    Z. Phys. B  47 (1982) 365.

\bibitem{fes2d}
	  A. Vespignani, R. Dickman, M. A. Mu\~noz, and S. Zapperi,
	  Phys. Rev. E 62 (2000) 4564.

\bibitem{manna1d}
	  R. Dickman, M. Alava, M. A. Mu\~noz, J. Peltola, 
	  A. Vespignani, and S. Zapperi,             
	  Phys Rev. E{\bf 64}, 056104 (2001). 

\bibitem{mnrst} 
	    R. Dickman, T. Tom\'e, and M. J. de Oliveira, 
	    Phys. Rev. E, in press; e-print: cond-mat/0203565. 

\bibitem{mancam}
	    R. Dickman, e-print: cond-mat/0204608.

\bibitem{rossi}
	  M. Rossi, R. Pastor-Satorras, and A. Vespignani, 
	  Phys. Rev. Lett.  85 (2000) 1803.

\bibitem{paczuski}
	  M. Paczuski, S. Maslov, and P. Bak,
	  Europhys. Lett. {\bf 27}, 97 (1994); {\bf 28}, 295 (1994).

\bibitem{socprl} 
	  A. Vespignani, R. Dickman, M. A. Mu\~noz, and Stefano Zapperi,  
	  Phys. Rev. Lett. {\bf 81}, 5676 (1998). 

\bibitem{granada} 
	   M. A. Mu\~{n}oz, R. Dickman, R. Pastor-Satorras, A. Vespignani, and S. Zapperi, 
	   in {\it Modeling Complex Systems}, 
	   Proceedings of the 6th Granada Seminar on Computational 
	   J. Marro and P. L. Garrido, eds., AIP Conference Proceedings v. 574 (2001); 
	   e-print: cond-mat/0011442. 

\bibitem{doi}
	  M. Doi,
	  J. Phys. A{\bf 9}, 1465; 1479 (1976).

\bibitem{jsp1}
	   R. Dickman,
	   J. Stat. Phys. {\bf 55}, 997 (1989). 

\bibitem{tdpprl}
	   R. Dickman and I. Jensen, 
	   Phys. Rev. Lett. {\bf 67}, 2391 (1991).

\bibitem{jsp2}
	 I. Jensen and R. Dickman,
	 J. Stat. Phys. {\bf  71}, 89 (1993). 

\bibitem{rsa} 
	 R. Dickman,  J.-S. Wang, and  I. Jensen, 
	 J. Chem. Phys. {\bf 94},  8252 (1991);       
	 M. J. de Oliveira, T. Tom\'e, and R. Dickman,
	 Phys. Rev. A {\bf 46}, 6294 (1992).
      
\bibitem{redner}
	    J. Zhuo, S. Redner, and H. Park,
	    J. Phys. A{\bf 26}, 4197 (1993). 

\bibitem{peliti85} 
      L. Peliti,  
      J. Physique {\bf 46}, 1469 (1985). 
 
\bibitem{pert} 
	    R. Dickman and R. R. Vidigal,
	    e-print: cond-mat/0205231. 

\bibitem{peliti86}
	 L. Peliti,     
	 J. Phys. A {\bf 19}, L365 (1986).

\bibitem{lee}
	 B. P. Lee,
	 J. Phys. A {\bf 27}, 2633 (1994).

\bibitem{wijlanddp}
	 F. van Wijland, K. Oerding, and H. J. Hilhorst,
	 Physica A{\bf 251}, 179 (1998).

\bibitem{manna} 
	S. S. Manna, 
	J. Phys. A {\bf 24}, L363 (1991). 
 
\bibitem{manna2} 
	 S. S. Manna,   
	 J. Stat. Phys. {\bf 59}, 509 (1990). 

\bibitem{rdunp}
	  R. Dickman, unpublished.

\bibitem{baram}
	 A. Baram and D. Kutasov,
	 J. Phys. A{\bf 22}, L251 (1989).
 
 
\end{thebibliography}
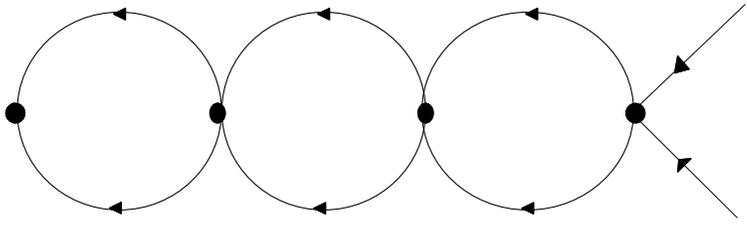
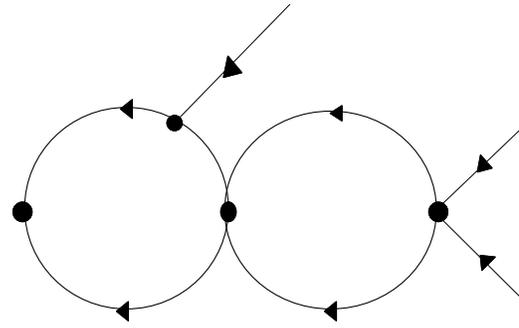
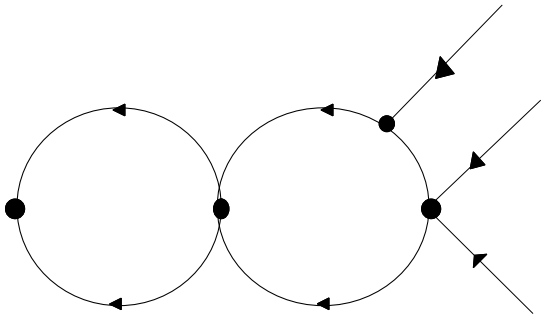
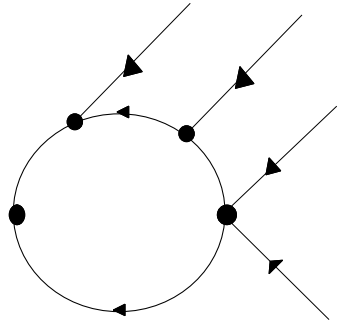
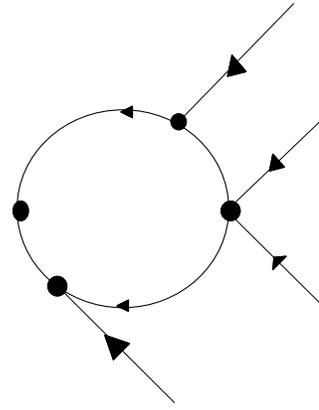
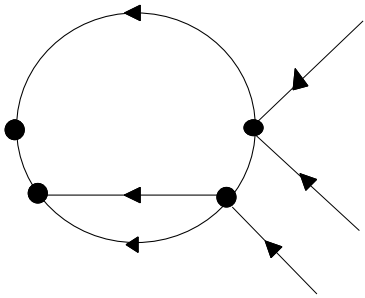
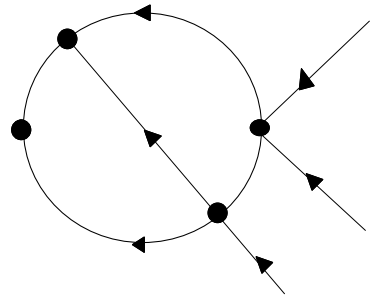
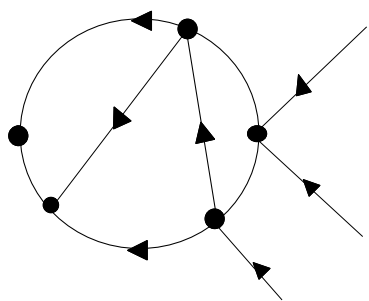
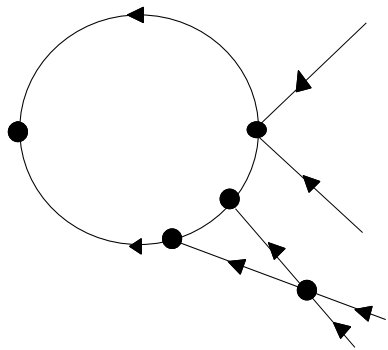
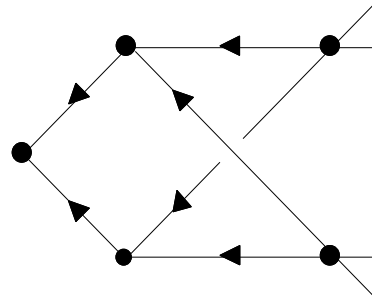
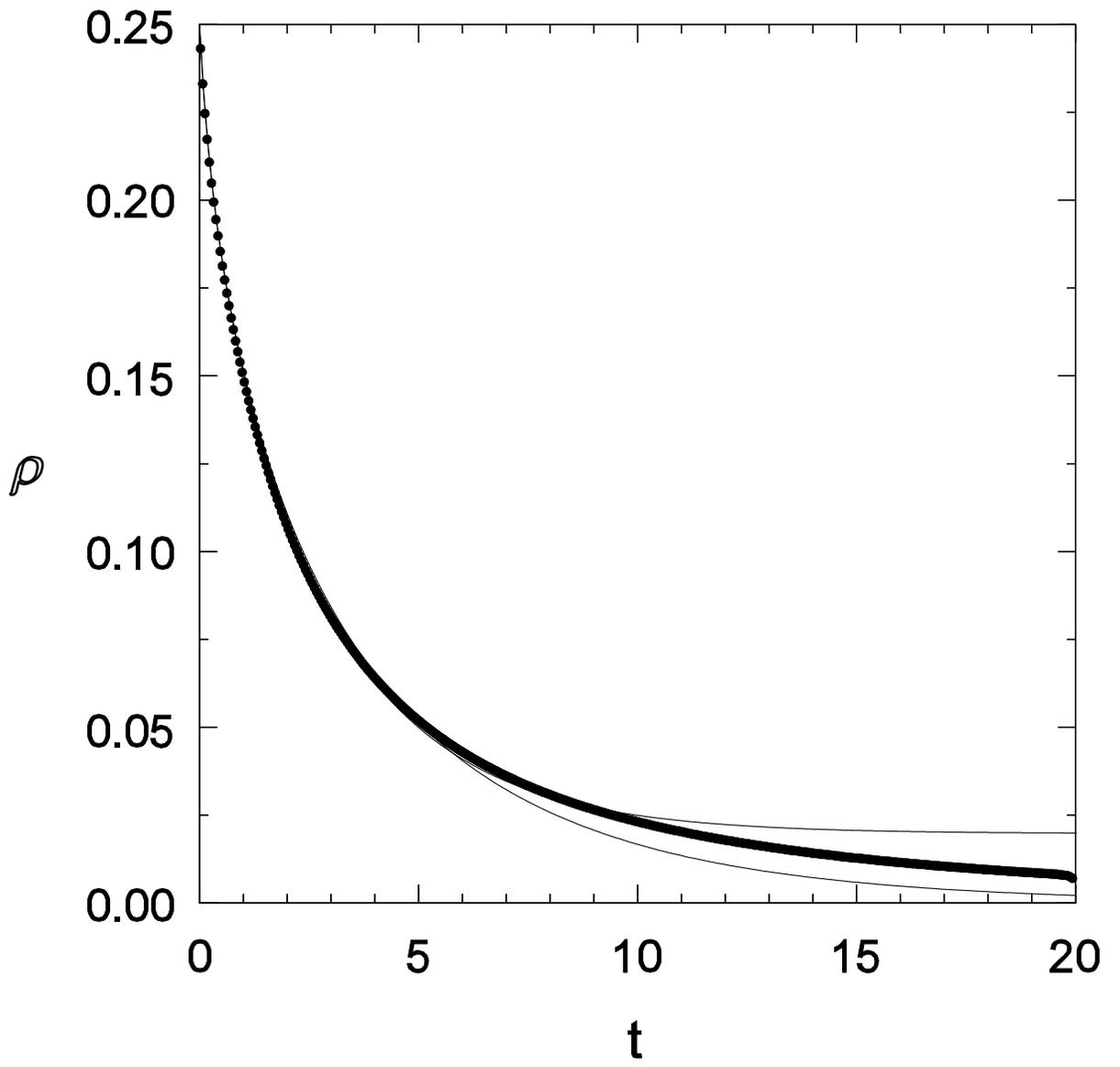
\end{document}